\documentclass[conference]{IEEEtran}
\usepackage{cite}
\usepackage{amsmath,amssymb,amsfonts}
\usepackage{algorithmic}
\usepackage{graphicx}
\usepackage{textcomp}
\usepackage{xcolor}
\def\BibTeX{{\rm B\kern-.05em{\sc i\kern-.025em b}\kern-.08em
    T\kern-.1667em\lower.7ex\hbox{E}\kern-.125emX}}
\begin{document}

\title{Methodology for forecasting and optimization in IEEE-CIS 3rd Technical Challenge}

\author{\IEEEauthorblockN{Richard Bean}
\IEEEauthorblockA{
\textit{Centre for Energy Data Innovation} \\
\textit{School of Information Technology and Electrical Engineering} \\
\textit{University of Queensland}\\
Brisbane, Australia \\
r.bean1@uq.edu.au
}
}

\maketitle

\begin{abstract}

This report provides a description of the methodology I used in the IEEE-CIS 3rd Technical Challenge.\cite{tech} 

For the forecast, I used a quantile regression forest approach using the solar variables provided by the Bureau of Meterology of Australia (BOM) and many of the weather variables from the European Centre for Medium-Range Weather Forecasting (ECMWF). 

Groups of buildings and all of the solar instances were trained together as they were observed to be closely correlated over time. Other variables used included Fourier values based on hour of day and day of year, and binary variables for combinations of days of the week.

The start dates for the time series were carefully tuned based on phase 1 and cleaning and thresholding was used to reduce the observed error rate for each time series. 

For the optimization, a four-step approach was used using the forecast developed. First, a mixed-integer program (MIP) was solved for the recurring and recurring plus once-off activities, then each of these was extended using a mixed-integer quadratic program (MIQP).

The general strategy was chosen from one of two (``array'' from the ``array'' and ``tuples'' approaches) while the specific step improvement strategy was chosen from one of five (``no forced discharge'').

\end{abstract}



\section{Introduction}

My research background in combinatorics involves 0-1 integer programming to find solutions to Latin square and graph theory problems.~\cite{bean2005critical,bean2004size,bean2006latin,oeis}

In my energy consulting work, I used quadratic programs to develop bidding strategies to simulate the National Electricity Market in Australia as in \cite{bean2011calculation}.

While working at a smart solar inverter manufacturer company~\cite{bean2018using} I developed battery schedules using forecasts based on quantile regression forests, with weather forecasts taken from the Global Forecasting System.

As bike sharing demand forecasting uses virtually the same regressors as energy forecasting, in~\cite{bean2021does} I used Generalized Additive Models and ERA-5 weather data to forecast demand across 40 cities in sixteen countries.

This background explains why I entered this competition and the design decisions I made given the limited time the competition was active.

\section{Background}
\subsection{Forecast}
I began building the forecast around 10 September, using the Generalized Additive Model as seen in~\cite{pierrot2011short} and my bike-sharing demand work~\cite{bean2021does}. This was to develop an initial feel for how temperature and solar variables in the ECMWF (ERA-5) data set affected each building and solar installation, along with temporal variables (weekend, time of day, and day of year). 

From the very first entry on 10 September, I noticed that the buildings were very different in terms of load on weekend and public holidays, and that temperature and solar (leading and lagging) were the most critical predictor variables in all the models. 

I quickly switched to a random forest model as the focus of the competition was purely the lowest error rate, rather than explainability or visualization.

The Global Energy Forecasting Competitions of 2014 and 2017 (\hspace{1sp}\cite{hong2016probabilistic,hong2019global}) provided strong evidence that random forests were one of the most successful techniques in energy forecasting.

Since the publication of ~\cite{bean2018using} the ``ranger'' package~\cite{wright2015ranger} in R has provided multi-threaded random forests with an extension of options over those seen in the original ``quantregForest'' package~\cite{meinshausen2006quantile}.

The solar traces seemed to be genuine 15 minute readings, while Buildings 0 and 3 were series of 15 minute values repeated 4 times each. I gradually realized that Building 4 and Building 5 readings could not be predicted using weather or temporal variables and used the median value from Oct 2020 (i.e. ``manual optimisation''). All of these observations saved time in the prediction.

I thresholded values from Building 0 and 3 as some of them were clearly outliers.

A ``maximal'' approach was used; that is, for each time series, the training data was taken back as far as possible until the error rate started increasing.

\subsection{Optimization}

My background is strongest in 0-1 integer optimization rather than linear programming so I looked for something I could build off. I had also used AMPL for modelling in the past, but the Python interface of Gurobi is now so well integrated that using a modelling language became an unnecessary hassle.

Taheri (\hspace{1sp}\cite{taheri1,taheri2}) created instructional videos on how to use Gurobi for solving scheduling problems. She suggested two methods for building a mixed integer program to minimize the cost of a schedule: the ``array'' (i.e. 0-1 IP) and ``tuple'' approaches. I watched these videos around 16 September. Source code was available from the Gurobi website~\cite{gurobicode}.

\section{Methodology}
\subsection{Forecast}


The forecast code was built in R as this is the programming language I was most familiar with for forecasting.

The following list shows forecast pseudo-code for quantile regression forest model development. Reproducibility is ensured by running the same program for Phase 1 and Phase 2 with only the ``PHASE'' parameter changed (although different random seeds would perturb the output slightly).

\begin{itemize}
    \item Pick a time series or group of time series
    \item Perform adjustment:
    \begin{itemize}
        \item Adjust start and end dates of training data
        \item Perform thresholding (effective for Solar 0 and Solar 5 series)
        \item Add or subtract predictor variables (e.g. leading and lagging variables, BOM variables)
        \item Group solar or buildings together differently
        \item Adjust random forest parameters (number of trees and mtry)
    \end{itemize}
    \item Assess MASE of Phase 1; if a change has resulted in a lower MASE, then  retain the change, otherwise discard it
\end{itemize}

After performing feature selection for each building and setting the value of Building 4 to be 1 kW for the whole month of October, I achieved an error rate (MASE) of 0.6528 by 20 September, which remained better than any other forecast on the leaderboard for Phase 1. This required the selection of start months for the buildings and solar data.

By the end of Phase 1, I had lowered this to 0.6320 by incorporating median forecasting of the time series and adding in BOM solar data.

On 13 October, the individual Phase 1 time series became available so I investigated how the ``0.6320'' value was derived using the given data and R program.

\begin{figure}
\centering
	\begin{center}
 		\includegraphics[width=3.4in]{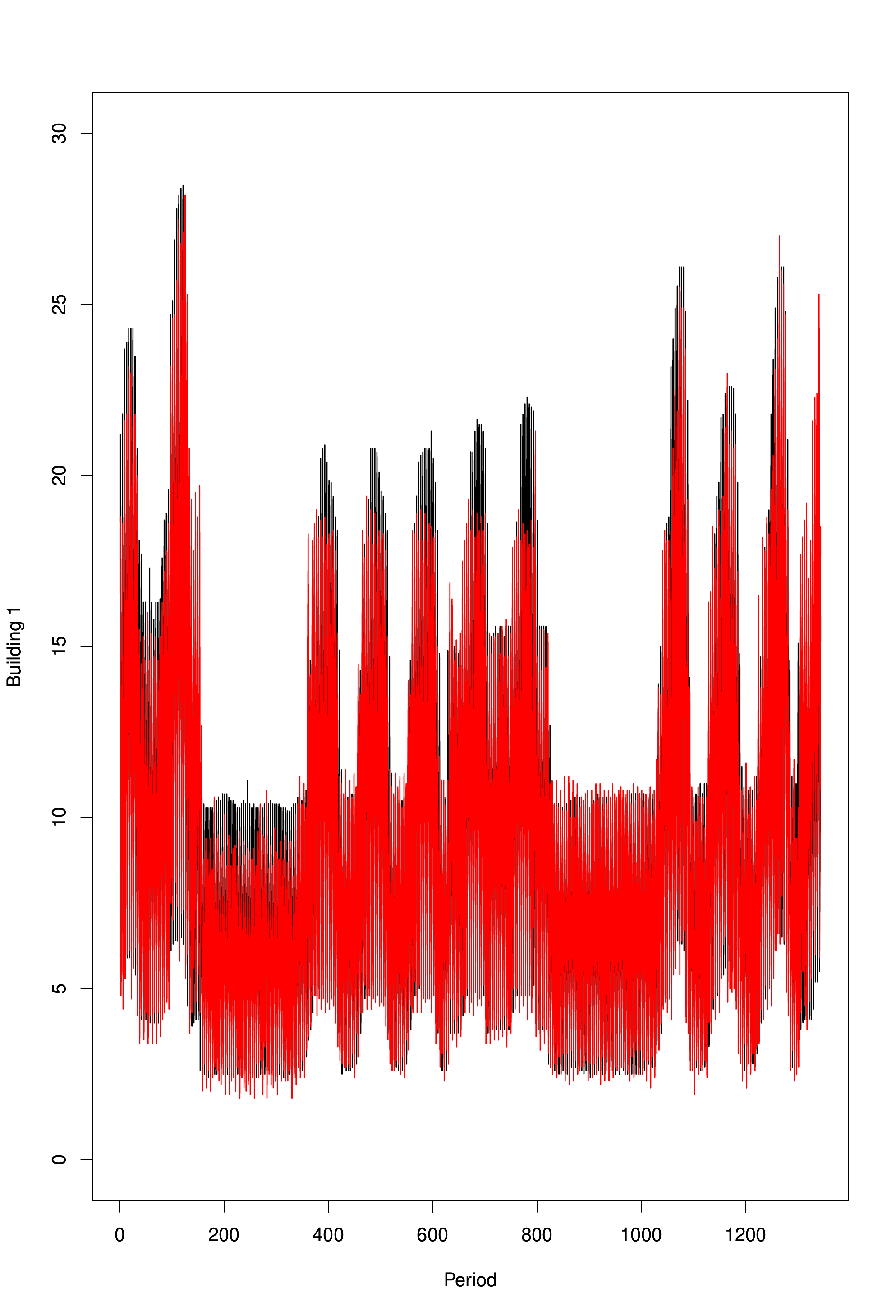}
 	\caption{\label{b1}Building 1 Forecast vs Actual  - Phase 1}
\end{center}
\end{figure}

\begin{figure}
\centering
	\begin{center}
 		\includegraphics[width=3.4in]{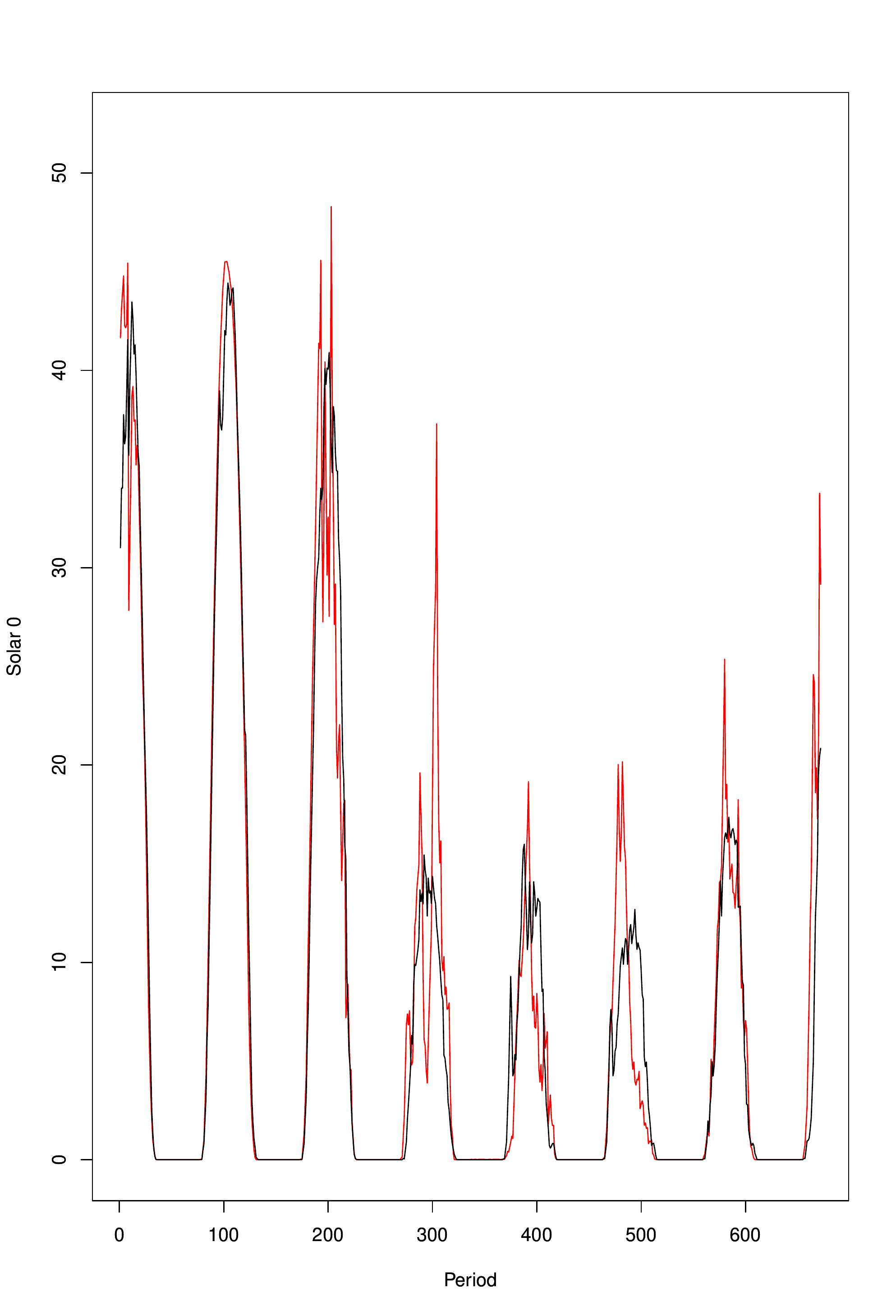}
 	\caption{\label{s0}Solar 0 Forecast vs Actual  - Phase 1}
\end{center}
\end{figure}

I then added the following improvements sequentially through experimentation as each change seemed reasonable and I thought each change would improve the error rate for Phase 2 (November 2020) as well.
\begin{itemize}
    \item Added cloud cover variables ${\pm}$ 3 hours; MASE 0.6243, 16 October; the effect may be seen in Fig~\ref{s0} with the predicted solar (black line) beginning to closely match the actual solar (red line) during the day
    \item Solar data from beginning of 2020 instead of from day 142; MASE 0.6063, 17 October
    \item Selected start month (0-8) for each of four building series from 2020, added all possible weather variables, set building 5 equal to 19 kW; MASE 0.5685, 18 October
    \item Fixed up Solar5 data by filtering out values less than 0.05 kW; MASE 0.5387, 24 October
    \item Noticed that forecasting Solar0 and Solar5 as linear combinations of the other Solar variables was working better than my actual Solar0/5 prediction
    \item Noticed that some pairs of solar series were much more highly correlated than other pairs, and buildings 3/6 were also highly correlated
    \item Tested exponential decay ideas, which were unsuccessful
    \item Trained all solar and building data together after seeing the Smyl and Hua paper~\cite{smyl2019machine} and recalling the competition abstract ``the opportunity for cross-learning across time series on two different prediction problems (energy demand and solar production).'' (MASE 0.5220, 30 October). A building forecast may be seen in Fig~\ref{b1}; the weekend/weekday pattern is clear.
    \item Fixed up Solar0 data by same filtering as for Solar5 (MASE 0.5207, 31 October)
    \item Added in separate binary variables for each day of the week (MASE 0.5166, 2 November)
\end{itemize}

\subsection{Optimization}
The optimization code was written in Python as Gurobi integrates much better with Python than with R. Thus, the key input files for the Python optimization were the vectors of prices for the month plus the vector of ``net'' load (that is, sum of building load minus sum of solar generation) for the month; 2,976 values for October 2020 (Phase 1) and 2,880 values for November 2020 (Phase 2).

After examining the Phase 2 instances, I decided to try to include all the once-off activities only in ``peak'' periods as this was much easier in MIP terms and the sum of the penalties for scheduling the once-off activities in ``off-peak'' exceeded the benefits for scheduling them in ``peak'' in every instance.

\begin{itemize}
    \item Develop a MIP for each of 5 small and 5 large instances - minimize the recurring load over all peak time periods
    \item Extend the solution of each MIP (Phase 2) to include all once-off activities in peak
    \item For each of 10 instances, solve the MIQP: attempt to add batteries and shift activities using the ``no forced discharge'' approach described below; consider all intermediate solutions
    \item Perform the same task for the ``recurring plus once-off'' solutions found
    \item Assess the cost using the objective function; if the ``recurring plus once-off'' cost for an instance is lower than the ``recurring only'' cost choose that solution, otherwise choose the ``recurring only'' solution
\end{itemize}

\begin{figure}
\centering
	\begin{center}
 		\includegraphics[width=3.5in]{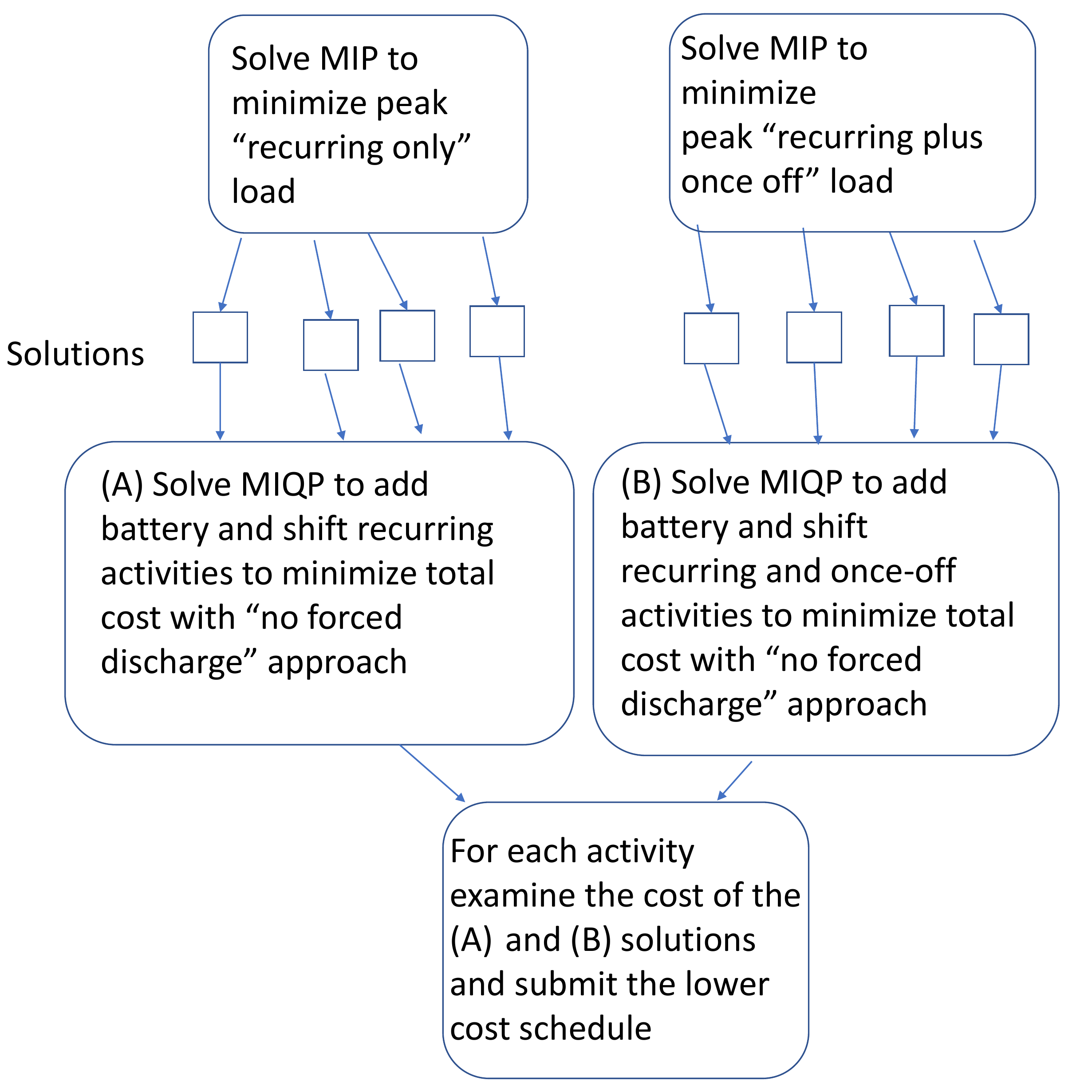}
 	\caption{\label{bp2}Phase 2 Optimization Solution Approach}
\end{center}
\end{figure}

The approach is shown in Fig~\ref{bp2}.

\section{Experiments}

The ``Optim Eval'' Java code was used to verify that the objective function cost of the final MIQP, solved in Gurobi, was aligned with the calculation used for the leaderboard and evaluation.



After attempting to solve the `small' and `large' instances as mixed-integer quadratic programs (MIQPs) with the quadratic term in the objective function, I soon realized that the cost of electricity from the pool prices (wholesale AEMO price) varied little between solutions, and the best solution would be much more easily found by solving a MIP (mixed integer program). That is, first minimize the recurring load over all peak time periods for the set of small and large instances, store all the intermediate solutions, and then attempt to solve the MIQP (incorporating the peak quadratic term in the objective function) by allowing Gurobi to check if moving activities around from the intermediate solutions could decrease the cost. Attempts to add constraints to the MIP to bias the solution away from weekdays with a higher average price were unsuccessful.

The organizers envisaged that forecast skill would have more effect in Phase 2 than Phase 1, but judging by the final Phase 2 leaderboard, it seems that there was little correlation. Perhaps having a commercial solver such as Gurobi and access to high-performance computing facilities were more important factors. In contrast, the forecasting task could be performed on a single computer in minutes.

It took me until 22 September to find an adjusted schedule which met the validity requirements of the precedences and room limits.

The Gurobi solver was the only one available on the University of Queensland HPC system and it performs very well in terms of solution speed (\hspace{1sp}\cite{hans,matt}).

There are parallels here to my past work in developing scheduling algorithms for home battery/inverter combinations~\cite{bean2018using}.

In that work, a key design decision was never to charge in peak hours and assess cost of different battery scheduling approaches over 83 inverters. The approaches included PV persistence, PV and Load persistence, Load persistence, quantiles of 50/50 and 60/40 for the PV and Load, and persistence of the last hour.

In the current work, I considered five approaches:

\begin{itemize}

    \item {\it Conservative} is just choosing the lowest recurring load and lowest recurring + once off load and evaluating cost using a naive or flat forecast. This was probably the winning approach for cost in Phase 1, as some competitiors had winning results with no forecast, or a poor forecast, but seemed pointless to me as the organizers said quality of forecast should contribute to results in phase 2.

    \item {\it Forced discharge} forbids any charging in peak hours, and forces at least one of the two batteries to be discharging in every peak period. This was thought to avoid nasty surprises in the peak load as in phase 1 one of the actual observed values (period 2702 of 2976) was ~260 kW above my final forecast (i.e. forecast with 0.5166 MASE). However, although values drop randomly in and out of the building data, I hoped that there were no ``outliers'' in phase 2 as promised (although this ``outlier'' comment from the competition organizers probably referred to the repeated 1744.1 kW values in the Building 0 trace - periods 1710 to 1713 of 2976).

    \item {\it No forced discharge} forbids any charging in peak hours, but the MIQP solver decides whether to discharge or do nothing in those hours.

    \item {\it Liberal} allows charging in peak, but the maximum of recurring + once off + charge effect for each period is limited to the maximum of recurring + once off load over all periods. This is to avoid nasty surprises when the solver thinks that a period has low underlying load and schedules a charge (due to a low price in that period) but then accidentally increases the maximum load over all periods, which can be very costly.

    \item {\it Very liberal} allows charging over peak and does not attempt to control the maximum of recurring + once off + charge effect. This would be the best approach if the forecast was perfect.
\end{itemize}

Each approach was assessed starting with the best Phase 1 solutions obtained (and the best forecast available for Phase 1).

The ``Optim Eval'' Java program provided by the competition organizers was used to calculate the cost of each approach. 

It was found the ``liberal'' and ``very liberal'' approaches resulted in the lowest objective function value for the MIQP; but the prices obtained were actually higher using the known load and solar values. Over the 10 sample problems, the total cost was lowest for ``no forced discharge'' (evaluated prices: \$396,264 for Forced, \$396,060 for No Forced Discharge) and so this approach was used for Phase 2. Ultimately only ``Large 2'' and ``Large 4'' solutions included once-off activities.
\vspace{-4pt}
\section*{Acknowledgment}

Thanks to Archie Chapman for mentioning the competition on the work Slack channel, and to David Green of UQ HPC for helping me with Gurobi on the Tinaroo cluster.
\vspace{-4pt}
\bibliographystyle{IEEEtran}
\bibliography{monash}
\end{document}